\documentclass[a4paper,11pt]{article}
\usepackage{amsthm}
\usepackage{amsmath}
\usepackage{amssymb}
\usepackage{hyperref}
\usepackage{latexsym}
\usepackage{enumerate}

\usepackage{color}
\usepackage{setspace}
\usepackage{comment}
\usepackage{graphicx}
\usepackage{float}

\usepackage{titletoc}
\dottedcontents{section}[1.2em]{}{1em}{1pc}

\usepackage{subfigure}
\usepackage{tikz}
\usetikzlibrary{decorations.pathmorphing}
\usetikzlibrary{decorations.pathreplacing}
\usepackage{caption}
\usepackage{pgf}
\usepackage{pgffor}
\usepackage[utf8]{inputenc}
\usepgfmodule{shapes}
\usepgfmodule{plot}
\usetikzlibrary{shapes.geometric}

\usetikzlibrary{decorations.markings}
\tikzset{->-/.style={decoration={
  markings,
  mark=at position .5 with {\arrow{>}}},postaction={decorate}}}
\newcommand{\ket}[1]{| #1 \rangle}
\title{Quantum Jump from Singularity to Outside of Black Hole}
\author{Furkan Semih Dündar\footnote{email: furkan.dundar1@ogr.sakarya.edu.tr} $\,^{a}$ and Kamal Hajian\footnote{email: kamalhajian@ipm.ir}$\,\,^{,b,c}$\\ 
\footnotesize $^a$\textit{Physics and Mathematics Departments, Sakarya University, 54050, Sakarya, Turkey}\\
\footnotesize $^b$\textit{School of Physics, Institute for Research in Fundamental
Sciences, P.O. Box 19395-5531, Tehran, Iran}\\
\footnotesize $^c$\textit{Department of Physics, Sharif University of Technology, P.O. Box 11365-8639, Tehran, Iran}
}
\date{}
\begin{document}

\maketitle
%--------------------------------------------------------------------------------------------------------------------------------------
\begin{abstract}
Considering the role of black hole singularity in quantum evolution, a resolution to the firewall paradox is presented. It is emphasized that if an observer has the singularity as a part of his spacetime, then the semi-classical evolution  would be non-unitary as viewed by him. Specifically, a free-falling observer inside the black hole would have a Hilbert space with non-unitary evolution; a quantum jump for particles encountering the singularity to outside of the horizon as late Hawking radiations. The non-unitarity in the jump resembles the one in collapse of wave function, but preserves entanglements. Accordingly, we elaborate the first postulate of black hole complementarity: freely falling observers who pass through the event horizon would have non-unitary evolution, while it does not have physically measurable effects for them. Besides, no information would be lost in the singularity. Taking the modified picture into account, the firewall paradox can be resolved, respecting No Drama. A by-product of our modification is that roughly half of the entropy of the black hole is released close to the end of evaporation in the shape of very hot Hawking radiation.  
\end{abstract}

%--------------------------------------------------------------------------------------------------------------------------------------
\setlength{\baselineskip}{1 \baselineskip}
%--------------------------------------------------------------------------------------------------------------------------------------
%------------------------------------------------------------------------------------------------------------------------------------------------------------------------------------------------
%------------------------------------------------------------------------------------------------------------------------------------------------------------------------------------------------

\begin{flushright}\vspace{-14cm}
{\small IPM/P-2015/013 \\
\today }\end{flushright}
\vspace{13cm}

{\tableofcontents}
\section{Introduction}

Since the introduction of evaporation process into the evolution of black holes in 1975 \cite{Hawking:1976rt}, the information paradox has been discussed {as an important and challenging problem} in the physics community. The essence of the argument is that while black holes absorb information in the form of matter or light, they leave a thermal density matrix when they evaporate. The vapor released by the black hole does not contain any fine quantum correlation found in the infalling matter or its quantum state. Hence, {it was} argued that black holes destroy information \cite{Hawking:1976ra} violating one of the widely held beliefs in physics. For a recent review of the information paradox, readers may see {refs.} \cite{Mathur:2009hf,Bultinck:2013em}.

A {well-known} resolution to the information paradox has been proposed in 1993 which is called the black hole complementarity \cite{Susskind:1993if}. It emphasizes the difference between the points of view of an observer remaining outside the black hole and an observer falling into the black hole. It brings the concept of a stretched horizon, a timelike surface just outside the event horizon. Whatever falls into the black hole is said to be absorbed by the stretched horizon. In terms of the outside observer, black hole is completely described by the stretched horizon degrees of freedom. The ``inside" of the black hole is a redundant term. However in terms of an infalling observer the stretched horizon does not exist, but inside of the black hole exist. These two views are complementary to each other. This is black hole complementarity. It argues that the following postulates produce a compatible unitary description of physics as seen by a an outside observer \cite{Susskind:1993if}:

\begin{itemize}

\item[] \textbf{Postulate 1:} 
 The process of formation and evaporation of a black hole, as viewed by a distant observer, can be described entirely within the context of standard quantum theory. In particular there exists a unitary S-matrix which describes the evolution from infalling matter to outgoing Hawking-like radiation.

\item[] \textbf{Postulate 2:} Outside the stretched horizon of a massive black hole, physics can be described to good approximation by a set of semi-classical field equations.

\item[] \textbf{Postulate 3:} To a distant observer, a black hole appears to be a quantum system with discrete energy levels. The dimension of the subspace of states describe a black hole of mass M is the exponential of the Bekenstein entropy $S(M)$.

\end{itemize}
In short, the black hole complementarity brings back the unitariness to the Hilbert space of an outside observer, in effect solving the information paradox. Although in \cite{Susskind:1993if} it is not emphasized explicitly as an additional postulate, the equivalence principle is assumed to hold as well. It is taken as the fourth postulate \cite{Almheiri:2012rt}, sometimes called \emph{No Drama};

\begin{itemize}

\item[] \textbf{Postulate 4:} A freely falling observer experiences nothing out of the ordinary when crossing the horizon.

\end{itemize}

In 2012 an argument has been provided \cite{Almheiri:2012rt,Almheiri:2013hfa} to show that the above postulates are not compatible with each other if one studies the physics as viewed by a freely falling observer carefully. It results in a paradox: the \emph{firewall paradox}. We will tell the paradox briefly in the next section, however the reader may see ref. \cite{Dundar:2014gma} for a recent review, and ref. \cite{Braunstein:2009my} for a precursor of the concept. {Then, in sections \ref{Our proposal}-\ref{sec hot explosion} we will present our proposal, describe how it resolves the paradox, provide some arguments supporting it, and investigate its properties and implications.}

\section{A Brief Summary of the Firewall Paradox}

In this section we will briefly tell the firewall paradox using the strong sub-additivity argument. In ref. \cite{Page:1993df} {it was} found that when the system is pure as a whole, any subsystem smaller than the half of it is almost maximally entangled with the rest, on average. Later this argument {was applied} to evaporating black holes in \cite{Page:1993wv}. {It was} found that if black hole radiation is unitary, after a certain time $t_{_\mathrm P}$, called as the ``Page time''  the entanglement entropy of the radiation begins to decrease. The radiation begins to be purified by the late Hawking radiation and in the end it becomes a pure state when the evaporation takes place completely.

\begin{figure}[b!]
    \vspace*{0.8cm}
\centering
\captionsetup{width=.8\textwidth}	
\begin{tikzpicture}[scale = 2]
    \draw (0,0) circle (1);
    \draw (0,1) node [anchor = south] {Event horizon};
    \tikzset{snake it/.style={decorate, decoration=snake,segment length=2mm}}
    \begin{scope}[scale=1]
    \draw[snake it] (0.9,0) -- (0.5,0);
    \draw[very thick,->,>=stealth] (0.47,0) to (0.45,0);
    %\draw (0.6,-0.01) node [anchor = east] {\scriptsize\raise1.25pt\hbox{$\blacktriangleleft$}};
    \end{scope}
    \begin{scope}[scale=1]
    \draw[snake it] (1.1,0) -- (1.5,0);
    \draw[very thick,->,>=stealth] (1.53,0) to (1.55,0);
    \end{scope}
    \begin{scope}[scale=1]
    \draw[snake it] (2.5,0) -- (2.9,0);
    \draw[very thick,->,>=stealth] (2.93,0) to (2.95,0);
    \end{scope}    
    \draw[fill=black] (0.9,0) circle (0.02) node [anchor = south] {$A$};
    \draw[fill=black] (1.1,0) circle (0.02) node [anchor = south] {$B$};
    \draw[fill=black] (2.5,0) circle (0.02) node [anchor = south] {$R$};
\end{tikzpicture}

\caption{\footnotesize For an old black hole, $B$ and $R$ are maximally entangled because of Postulate 1. $A$ and $B$ are maximally entangled because of Postulate 4. However, this series of mutual entanglements is excluded by the strong sub-additivity of entropy. Hence the firewall paradox.}
\label{fig:ABR}
\end{figure}
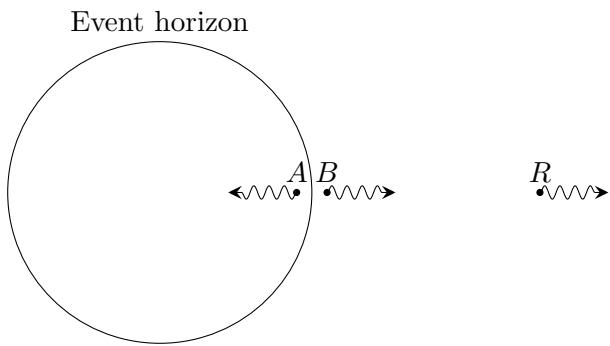

In ref. \cite{Almheiri:2012rt}, authors (AMPS for short) consider a black hole past the Page time: an \emph{old} black hole. The radiation it has already emitted is called the early radiation. Now, suppose it radiates a Hawking quantum, which we name as particle $B$. Because these quanta are created in pairs, it has an infalling partner towards the singularity, particle $A$. At the Page time, the black hole has half of its initial entropy \cite{Page:1993wv}. So it is the smaller subsystem. The particle it radiates into the space, particle $B$, is maximally entangled with a subsystem of the early radiation, dubbed $R$ (See Figure \ref{fig:ABR}). This is required by the unitarity of quantum mechanics: Postulate 1. On the other hand, particle $A$ and particle $B$ must be maximally entangled as well, because the state they create together must be vacuum with respect to infalling observers. This is required by the equivalence principle: Postulate 4. The mutual maximal entanglements between $A$--$B$ and $B$--$R$ are not allowed because of the strong sub-additivity of entropy. Hence we obtain a paradox.

At some later time, particle $B$ becomes a well defined Hawking particle, say, well outside the event horizon. Then, using the semi-classical approximation, Postulate 2, we can trace its evolution backwards in time until near the event horizon. Because of the huge blue shift factor it will have a huge energy. Hence, ref. \cite{Almheiri:2012rt} suggests that there is a wall of fire near the event horizon, violating the equivalence principle according to which it should be in vacuum state because it is not a high curvature regime. Therefore, ref. \cite{Almheiri:2012rt} argues that black hole complementarity is inconsistent in itself, resulting in the firewall paradox.

\section{Singularity and Non-unitarity Inside Black Holes}\label{Our proposal}
{Before explaining our resolution to the firewall paradox, we present an alternative way yielding this paradox which helps enlightening the road towards the resolution. Let us consider an old black hole, with the early and late radiations $R$ and $B$ similar to what AMPS had focused on. As it was mentioned in the previous section, by the Page-AMPS argument the early and late radiations $R$ and $B$ have to be maximally entangled, see Figure~\ref{fig:BR}.
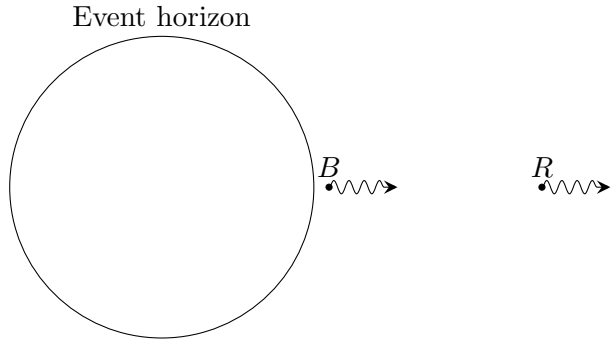
\begin{figure}[H]
    \centering
\vspace*{0.2cm}
\captionsetup{width=.82\textwidth}	
\begin{tikzpicture}[scale = 2]
    \draw (0,0) circle (1);
    \draw (0,1) node [anchor = south] {Event horizon};
    \tikzset{snake it/.style={decorate, decoration=snake,segment length=2mm}}
    \begin{scope}[scale=1]
    \draw[snake it] (1.1,0) -- (1.5,0);
    \draw[very thick,->,>=stealth] (1.53,0) to (1.55,0);
    \end{scope}
    \begin{scope}[scale=1]
    \draw[snake it] (2.5,0) -- (2.9,0);
    \draw[very thick,->,>=stealth] (2.93,0) to (2.95,0);
    \end{scope}    
    \draw[fill=black] (1.1,0) circle (0.02) node [anchor = south] {$B$};
    \draw[fill=black] (2.5,0) circle (0.02) node [anchor = south] {$R$};    
\end{tikzpicture}

\caption{\footnotesize For an old black hole, $B$ and $R$ are maximally entangled because of Postulate 1.}
\label{fig:BR}
\end{figure}
\noindent Now, instead of requesting No Drama at the late time when $B$ evaporates, we request No Drama at the time of evaporation of $R$, \emph{i.e.} at early times. At early times, No Drama has to be incarnated by pair production process. As a result, $R$ and its  interior partner, calling it $Q$, should be in a maximally entangled pure state. Now, by Postulate 1 unitary evolutions would not disentangle this maximally entangled pure state. Therefore, $Q$ and $R$ would remain maximally entangled at all times, specifically at the late time when $B$ evaporates. Consequently, at that late time $R$ would be maximally entangled with both $Q$ \emph{and} $B$. But the mutual maximal entanglements of $Q$--$R$ and $B$--$R$ are not allowed because of the strong sub-additivity of entropy. Therefore, one would encounter contradiction; the firewall paradox. Advantage of this version of the paradox is that it suggests the resolution too: if $Q$ could be identified with $B$ using a consistent mechanism, then this version of paradox would be resolved. For sure, any proposed mechanism has to resolve other versions, \emph{i.e.} the AMPS ``monogamy" and "blue shifted state" versions described in the previous section. In the remaining of the paper, we will try to construct such mechanism.}

Our proposal to resolve the paradox is based on emphasising the role of singularity in the evaporation process. The idea is based on an expectation: if an observer has the singularity as a part of his spacetime, time evolution can not be unitary at the semi-classical level as viewed by him. To be confident about this expectation, it would be enough to consider a particle which encounters the singularity and disappears there\footnote{{It has been shown recently that Schwarzschild geometry is not $C^0$ extendible \cite{Sbierski:2015nta}. So, the geodesic of a classical particle encountering the singularity can not be extended anymore.}}. We elaborate the first postulate of black hole complementarity, which brings this role of singularity to the scene. The key step toward the modification is to make a distinction between the freely falling observers outside the event horizon and the ones inside. 

Remembering the stretched horizon, the basic feature of an outside observer is that he has not quantum states around the black hole singularity as a part of the Hilbert space. Similarly we request that the Hilbert space of a freely falling observer \emph{outside} the event horizon be composed of the outside region of black hole plus the stretched horizon. So his Hilbert space would not include states around the singularity of the black hole, and his evolution would be unitary. On the other hand, the Hilbert space of a freely falling observer \emph{inside} of the horizon can be affected by singularity as a part of his/her spacetime. From the point of view of that observer the evolution would \emph{not} be unitary, in the semi-classical regime. In brief, we expect a transition from unitarity to non-unitarity for an infalling observer passing the event horizon. This idea is the origin of elaborating the first postulate as follows:

\begin{itemize}

\item[]{\textbf{Postulate 1$'$:}}
The process of formation and evaporation of a black hole, as viewed by an outside observer, can be described entirely within the context of standard quantum theory. In particular there exists a unitary S-matrix which describes the evolution from infalling matter to outgoing Hawking-like radiation.

\item[]Time evolution is non-unitary for observers inside the black hole. The non-unitariness is always in a way that does not have any measurable effect for them, and no information is lost in the overall picture.
\end{itemize}

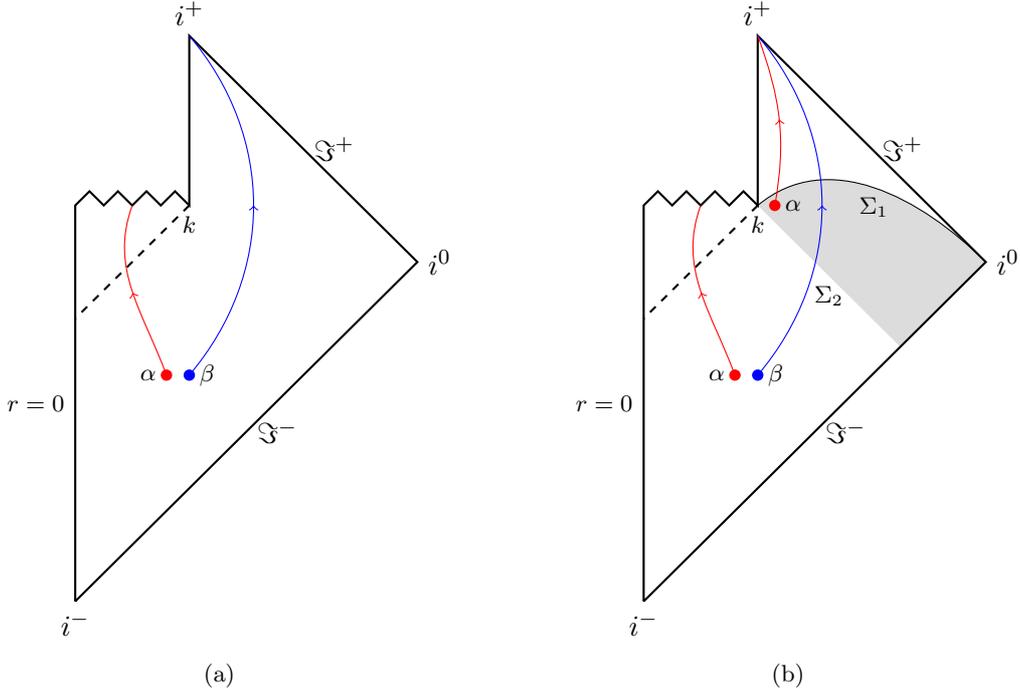
\begin{figure}[t!]
  	\captionsetup{width=.8\textwidth}	
	\centering
	\subfigure[]{
	\begin{tikzpicture}[scale=1.5]
    \draw[thick](0,0)node[below] {$i^-$} -- node[midway, left] {\footnotesize $r=0$} (0,3.5)-- (0.125,3.625)-- (0.25,3.5)-- (0.375,3.625)-- (0.5,3.5)-- (0.625,3.625)-- (0.75,3.5)-- (0.875,3.625)-- (1,3.5)--(1,5)node[above] {$i^+$} -- node[midway, right] {$\Im^+$} (3,3)node[right] {$i^0$}-- node[midway, right] {$\Im^-$} (0,0); 
    \draw[thick,dashed](1,3.5) -- (0,2.5);
    \fill[blue] (1,2) circle (0.05cm);
    \draw (1,2) node[right] {\footnotesize $\beta$};
    \draw (1,3.5) node[below] {\footnotesize $k$};
    \draw[blue,->-] (1,2) to [out=50,in=310] (1,5); 
    \fill[red] (0.8,2) circle (0.05cm);
    \draw[red,->-] (0.8,2) to [out=110,in=250] (0.5,3.5);
    \draw (0.8,2) node[left] {\footnotesize $\alpha$};
	\end{tikzpicture}}\hspace{1cm}
	\definecolor{light gray}{RGB}{220,220,220}
	\subfigure[]{
    \begin{tikzpicture}[scale=1.5]
    \fill[even odd rule,light gray](1,3.5) to [out=40,in=135] node[black,midway,below] {\footnotesize $\Sigma_1$} (3,3) to (2.25,2.25) to node[black,midway,below] {\footnotesize $\Sigma_2$} (1,3.5);
    \draw[black] (1,3.5) to [out=40,in=135] (3,3); 
    \draw[thick](0,0)node[below] {$i^-$} -- node[midway, left] {\footnotesize $r=0$} (0,3.5)-- (0.125,3.625)-- (0.25,3.5)-- (0.375,3.625)-- (0.5,3.5)-- (0.625,3.625)-- (0.75,3.5)-- (0.875,3.625)-- (1,3.5)--(1,5)node[above] {$i^+$} -- node[midway, right] {$\Im^+$} (3,3)node[right] {$i^0$}-- node[midway, right] {$\Im^-$} (0,0); 
        \draw[thick,dashed](1,3.5) -- (0,2.5);
    \fill[blue] (1,2) circle (0.05cm);
    \draw (1,2) node[right] {\footnotesize $\beta$};
    \draw (1,3.5) node[below] {\footnotesize $k$};
    \draw[blue,->-] (1,2) to [out=50,in=310] (1,5); 
    \fill[red] (0.8,2) circle (0.05cm);
    \draw[red,->-] (0.8,2) to [out=110,in=250] (0.5,3.5);
    \draw (0.8,2) node[left] {\footnotesize $\alpha$};
    \fill[red] (1.15,3.5) circle (0.05cm);
    \draw (1.15,3.5) node[right] {\footnotesize $\alpha$};    
    \draw[red,->-] (1.15,3.5) to [out=80,in=290] (1,5);
    	\end{tikzpicture}\label{fig non-unitary frefo 2}}
	\caption{\footnotesize \textcolor{black}{Evolution of two entangled particles $\alpha$ and $\beta$. One is dropped towards the black hole and the other one remains outside, as viewed by a freely falling observer inside the black hole. (a) Without the Postulate {1$'$}: the particle $\alpha$ is destroyed at singularity, resulting in non-unitary evolution and information loss. (b) With the Postulate {1$'$}: the particle $\alpha$ disappears at singularity and re-appears outside the black hole, as a part of the evaporation process. Although this process is still non-unitary, it prevents information loss. The gray region shows the allowed part of the spacetime for its re-appearance in this jump. In addition, by complementarity, $\alpha$ has to jump into a specific region of that gray region. It will be studied and determined in Section \ref{sec hot explosion}}}\label{fig non-unitary frefo}
\end{figure}

The scenario is clarified in Figure \ref{fig non-unitary frefo} schematically. The figure shows Penrose diagram for a spherically symmetric black hole formed by collapsing matter. In addition, there are two entangled particles $\alpha$ and $\beta$ outside the black hole. From the point of view of an infalling observer inside the black hole, the former drops to the black hole and is ``destroyed" at singularity, while the latter remains outside. {The physics as seen by that infalling observer is non-unitary}: a result of semi-classical quantum theory and the presence of the singularity. If one assumes the destruction of particle $\alpha$ as its final fate, then the information would be lost. But by the Postulate {1$'$}, particle $\alpha$ disappears at the singularity, and re-appears  outside of the black hole as a part of the evaporation process. It would remain entangled with the particle $\beta$, preventing loss of information. 

A natural question arises about the position and time which $\alpha$ re-appears outside of the evaporating black hole. The gray region in Figure \ref{fig non-unitary frefo 2} shows a typical spacetime region which is allowed by causality. Besides, because of complementarity, there are some spatial and temporal bounds in that region. The reasoning is as following. 

\begin{itemize}
\item[\textbf{(I)}] \emph{The particle should be a part of Hawking radiation. Hance, it has to re-appear before complete evaporation of the black hole.}

In Figure \ref{fig non-unitary frefo 2}, a typical spacelike surface which determines complete evaporation of the black hole as seen by an outside observer is depicted by $\Sigma_1$. It has to connect $i^0$ to $k$, and be spacelike everywhere. By \textbf{(I)}, the particle has to appear in the region below the $\Sigma_1$.\\
\item[\textbf{(II)}] \emph{The particle should not causally affect its history.}

The worldline of a particle which eventually hits the singularity has to be timelike or lightlike, and in the region below or on the $\Sigma_2$. Therefore, by \textbf{(II)}, the particle has to re-appear in the region above the $\Sigma_2$. \\
\item[\textbf{(III)}] \emph{By complementarity, different observers  should be in agreement on observables outside the horizon.}
 
{From the point of view of an outside standing observer, particles in Hawking radiation (either early or late radiations) originate from usual evaporation of the stretched horizon. On the other hand, an infalling observer inside the black hole would categorize particles in the Hawking radiation outside, to two parts: 1) ones which genuinely have been created outside via pair production, 2) ones created outside by the quantum jump from singularity inside. In a distance far enough the horizon, the two complementary pictures should coincide.  This request puts more constraints on the spacetime region where jumps can occur. We postpone this analysis to Section \ref{sec hot explosion}.}
\end{itemize}

In the next section, according to the modified version of complementarity we expressed here, resolution of the firewall paradox is presented. Then, after listing some remarks in support of the proposal, an argument is  provided investigating whether non-unitary evolution experienced by freely falling observers would have any measurable effects for them.  

\section{Resolution of the Firewall Paradox}\label{sec Firewall}  

So far people have put forward many interesting resolution attempts for the firewall paradox. One may see Harlow-Hayden conjecture \cite{Harlow:2013tf}, strong complementarity principle \cite{Ilgin:2013iba,Bousso:2012as}, ER=EPR \cite{Maldacena:2013xja}, fuzzball complementarity \cite{Mathur:2013gua}, Papadodimas-Raju proposal \cite{Papadodimas:2012aq,Harlow:2014yoa,Papadodimas:2013jku, Papadodimas:2013wnh}, extreme cosmic censorship conjecture \cite{Page:2013mqa}, icezones \cite{Hutchinson:2013kka}, shape dynamics approach \cite{Gomes:2013bbl}, inclusion of the back-reaction \cite{Mersini-Houghton:2014zka, Mersini-Houghton:2014cta}, considering quantum fluctuations of the collapsing shell \cite{Brustein:2013qma,Brustein:2013uoa}, balanced holography interpretation of black hole entropy \cite{Verlinde:2013uja,Verlinde:2013vja}, or considering non-violent non-locality \cite{Giddings:2012gc,Giddings:2013kcj}. 

In addition, there have been some recent inclinations towards studying physics inside the black holes, usually to ameliorate the information and/or firewall paradoxes. For example, considering the black hole interior as a highly quantum system \cite{Brustein:2015sma}, studying effects of quantum fluctuations inside the black hole \cite{Abedi:2015yga}, information recovery out of interior region through radiations from the collapsing matter \cite{Kawai:2015uya}, bouncing back of the collapsed matter called ``Planck stars" \cite{Rovelli:2014cta}, or considering black hole singularity as an erasure of information based on freely falling lattice discretization \cite{Lowe:2015eba}.

\textcolor{black}{According to our view, resolution to the paradox can be explained as follows. In the argument of AMPS, there is an explicit assumption about infalling observers. It is ``in order to prevent drama for these observers, any particle in Hawking radiation close but outside the horizon should be entangled with an interior partner". This assumption has its origin from the analogy of infalling/outstanding observers in black hole geometry and free/accelerating observers in Rindler spacetime. We argue that this analogy is problematic because of a crucial difference: the presence of singularity in black hole geometry. We relax their explicit assumption, by proposing a consistent role for the singularity in the Hilbert space of infalling observers who have singularity as a part of their accessible spacetime. As a result, in our scenario, there are some particles in Hawking radiation outside and close to the horizon which do not have any interior partner. They have jumped there after encountering the singularity inside the black hole.}

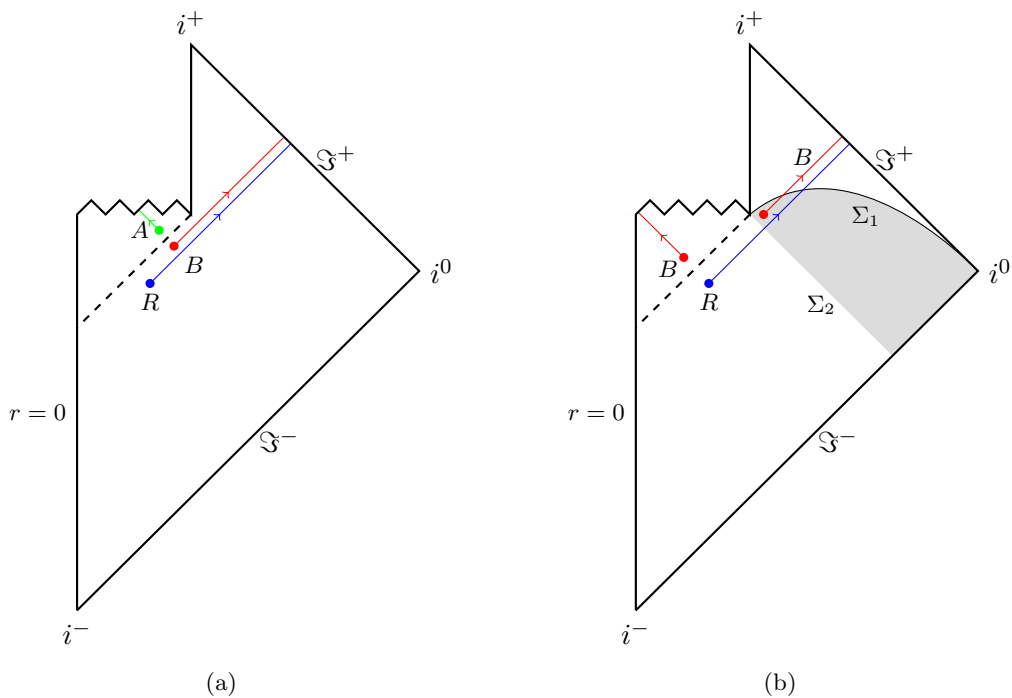
\begin{figure}[t!]
      %\vspace*{-0.7cm}
	\definecolor{light gray}{RGB}{220,220,220}
	\captionsetup{width=.8\textwidth}	
	\centering

	\subfigure[]{
	\begin{tikzpicture}[scale=1.5]
	
    \draw[thick,dashed](1,3.5) -- (0,2.5);
    \fill[blue,shift={(0.64cm,2.89cm)}] (0,0) circle (0.04cm);
    \draw[shift={(0.64cm,2.89cm)}] (0,0) node[below] {\footnotesize $R$};
    \draw[blue,shift={(0.64cm,2.89cm)},->-] (0,0) to (1.23,1.23);
     
    \fill[red,shift={(0.85cm,3.22cm)}] (0,0) circle (0.04cm);
    \draw[red,shift={(0.85cm,3.22cm)},->-] (0,0) to (0.96,0.96);    
    \draw[shift={(0.85cm,3.22cm)}](0,0) node[below right] {\footnotesize $B$};
    
    \fill[green,shift={(0.72cm,3.36cm)}] (0,0) circle (0.04cm);
    \draw[green,shift={(0.72cm,3.36cm)},->-] (0,0) to (-0.18,0.18);
    \draw[shift={(0.72cm,3.36cm)}](0,0) node[left] {\footnotesize $A$};
    
    \draw[thick](0,0)node[below] {$i^-$} -- node[midway, left] {\footnotesize $r=0$} (0,3.5)-- (0.125,3.625)-- (0.25,3.5)-- (0.375,3.625)-- (0.5,3.5)-- (0.625,3.625)-- (0.75,3.5)-- (0.875,3.625)-- (1,3.5)--(1,5)node[above] {$i^+$} -- node[midway, right] {$\Im^+$} (3,3)node[right] {$i^0$}-- node[midway, right] {$\Im^-$} (0,0); 
    	\end{tikzpicture}\label{fig Firewall paradox}}\hspace{1cm}
		\subfigure[]{
	\begin{tikzpicture}[scale=1.5]
\fill[even odd rule,light gray](1,3.5) to [out=40,in=135] node[black,midway,below] {\footnotesize $\Sigma_1$} (3,3) to (2.25,2.25) to node[black,midway,below] {\footnotesize $\Sigma_2$} (1,3.5);
    \draw[black] (1,3.5) to [out=40,in=135] (3,3);
    
    \fill[blue,shift={(0.64cm,2.89cm)}] (0,0) circle (0.04cm);
    \draw[shift={(0.64cm,2.89cm)}] (0,0) node[below] {\footnotesize $R$};
    \draw[blue,shift={(0.64cm,2.89cm)},->-] (0,0) to (1.23,1.23);
    
    \fill[red,shift={(1.12cm,3.50cm)}] (0,0) circle (0.04cm);
    \draw[red,shift={(1.12cm,3.50cm)},->-] (0,0) to node[midway,above,black]{\footnotesize $B$} (0.69,0.69);    
    \draw[shift={(1.12cm,3.50cm)}](0,0);
    
    \fill[red,shift={(0.42cm,3.12cm)}] (0,0) circle (0.04cm);
    \draw[red,shift={(0.42cm,3.12cm)},->-] (0,0) to (-0.4,0.4);
    \draw[shift={(0.46cm,3.18cm)}](0,0) node[below left] {\footnotesize ${B}$};    
    
    \draw[thick](0,0)node[below] {$i^-$} -- node[midway, left] {\footnotesize $r=0$} (0,3.5)-- (0.125,3.625)-- (0.25,3.5)-- (0.375,3.625)-- (0.5,3.5)-- (0.625,3.625)-- (0.75,3.5)-- (0.875,3.625)-- (1,3.5)--(1,5)node[above] {$i^+$} -- node[midway, right] {$\Im^+$} (3,3)node[right] {$i^0$}-- node[midway, right] {$\Im^-$} (0,0); 
    \draw[thick,dashed](1,3.5) -- (0,2.5);
	\end{tikzpicture}\label{fig Firewall resolution}}

	\caption{\footnotesize The firewall paradox and its resolution. (a) The firewall paradox: $R$ is a subset of early radiation, and is maximally entangled with the late radiation $B$ because of the unitarity of Hawking radiation process. According to the equivalence principle, $B$ has to be maximally entangled with a state $A$ inside the black hole. Being maximally entangled with two systems, $R$ and $A$, at the same time contradicts the strong sub-additivity of entanglement entropy. (b) Resolution: $B$ has been the inner partner of $R$. They have been entangled as pair particles, from the moment of the evaporation of $R$. They remain entangled during the evaporation process. $B$ inevitably encounters the singularity, disappears there and then re-appears outside of the black hole as the late radiation. Without contradicting the equivalence principle, there would not be any $A$; resolving the firewall paradox.}
\end{figure}

We have illustrated the evaporation process in AMPS scenario and ours in Figures \ref{fig Firewall paradox} and \ref{fig Firewall resolution} respectively. In \ref{fig Firewall resolution}, the late radiation $B$ has been the inner partner of the $R$ at the moment of emission of $R$. At the end of its journey inside of the black hole, it inevitably reaches to singularity, disappears there and re-appears as the late radiation outside of the black hole. It has been entangled with $R$ from the beginning, because of the equivalence principle. It will remain entangled with $R$ during its evolution, even after the complete evaporation of the black hole. Therefore there would not be any necessity to introduce the $A$. As a result, this process would be in agreement with monogamy of the entanglement, resolving the ``monogamy" version of the firewall paradox. In the appendix A, connection between non-unitariness and missing interior partners for $B$ modes, \textit{i.e.} the absence of $A$ modes is discussed more. 

It is also worth mentioning that it is impossible for any freely falling observer who is falling into the black hole, to encounter the $B$ radiation outside the black hole. This issue can be easily seen from the causal structure of Figure \ref{fig Firewall resolution}, in agreement with the Postulate {1$'$}. The worldline of any free-falling observer which would pass the event horizon has to be below $\Sigma_2$ in Figure \ref{fig Firewall resolution}. On the other hand, the $B$ radiation outside the black hole would jump into the region above $\Sigma_2$. So there is not possibility to encounter this radiation for that observer, outside the black hole. This argument resolves the ``blue shifted state" version of firewall paradox, \emph{i.e.} the infalling observer would not encounter any firewall on the horizon. 

{Concerning the version of the firewall paradox we presented at the beginning of Section \ref{Our proposal}, it is clear that the interior partner of $R$ has been identified with the late radiation $B$ which would be maximally entangled with $R$. So, this version of the firewall paradox is also resolved.}

\section{Remarks Supporting the Argument}
 In this section, we provide some remarks which support the argument presented. The reader might be interested to complete these remarks in order to provide a pros and cons list. 
\paragraph{No information loss:} In the black hole complementarity, although the assumed unitarity for a distant observer prevents information loss, but the question remains for infalling observers. For these observers, it has been assumed that the particles encountering the singularity are destroyed there. Hence, one might ask about conservation of information for them. Following our proposal, this question is answered clearly. There would not be any information loss for these observers too. Specifically, the jump does not disentangle entangled particles.
\paragraph{No baryon number violation:} Consider a baryon, \emph{e.g.} a proton, is sent to fall into a black hole. In the usual black hole complementarity, as viewed by infalling observers, that particle encounters the singularity and is destroyed there. It has been argued (see \emph{e.g.} section 9.3 in \cite{Susskind:2005js}) that in order to have complementary observation for a distant observers, the particle has to be thermalized completely on the strecthed horizon, perhaps by a decaying process.  Hence, for the distant observer, that particle disappears in the thermalization process, contradicting the stability of some baryons. In our proposed scenario, that particle is stable, and will constitute a part of evaporation process. 
\paragraph{No quantum xeroxing:} In quantum mechanics, it is impossible to have a unitary transformation (envisaged as a xerox machine) which duplicates any pure state $|\psi\rangle$ acted upon. In the usual black hole complementarity, a particle falling into the black hole passes the horizon as seen by an infalling observer, while is thermalized and evaporated as seen by a distant observer. One might be sceptic for quantum xeroxing, if the state outside would be reflected back towards the interior region of the black hole. It is because the infalling observer would have two copies of $|\psi\rangle$. To overcome this criticism, a proponent of black hole complementarity needs to provide physical arguments which prohibits that observer having access to the copied state (\emph{e.g.} see \cite{Susskind:2005js}). In our scenario, this problem is easily resolved by the impossibility of sending the jumped particles back to the interior, \emph{i.e.} the condition \textbf{(II)}.
\paragraph{Reproducing Page entanglement:} It has been argued that if one assumes the pair production at the horizon to be described by a pure maximally entangled state, then integrating out the interior partner results to entanglement entropy proportional to $\mathrm{N}\log 2$ in which $\mathrm{N}$ is the number of pairs created \cite{Mathur:2012jk}. Therefore, entanglement between the black hole and radiated particles outside would be monotonically increasing to the end of evaporation. On the other hand, Page analysis shows that the entanglement entropy between black hole and radiation has to decrease after the Page time \cite{Page:1993df,Page:1993wv}. So, we encounter a discrepancy \cite{Mathur:2012jk}. But, by the quantum jumps introduced in this paper the discrepancy is resolved; before the Page time, the early radiation is created by pair production which increases the entanglement entropy. After the Page time, the late radiation is produced by quantum jump (see Section \ref{sec hot explosion}). Leakage of interior partners to outside region through jumps decreases the entanglement entropy. It is simply because one should not trace out the jumped particles in order to find the entanglement entropy between the black hole and radiation. Notice that after the Page time, there should not be any pair production on the horizon, in order to prevent firewall paradox.
{\paragraph{Describing origin of the entanglement between early and late radiations:} Although Page-AMPS argument necessitates entanglement between early and late radiations, but its dynamical origin needs more explanation. Identifying interior partners of the early radiations with the late radiations through the quantum jump mechanism provides a natural dynamical explanation for that entanglement. The origin of the entanglement between early and late radiation would be their partnership in pairs created at the horizon.}

\section{Investigating Measurability of the Jumps}
One of the cornerstones of quantum mechanics is the unitarity of evolution. Nonetheless, there is an exceptional case which this celebrated axiom is not manifest; the collapse of the wave function in a measurement process. This seemingly non-unitariness is believed to be harmless to quantum physics because of its two main characteristics. Firstly, measurement always include a macroscopic measuring device. It is believed that if one takes that device into account, then the overall evolution might be unitary (\emph{e.g.} see \cite{joos2013decoherence,schlosshauer2007decoherence} and references therein). Secondly, that non-unitariness, although seems to have a non-local property, cannot be used to send measurable faster-than-light signals. {It is worth emphasizing that this approximation, replacing complicated measuring system by the collapse assumption, has resulted to major progresses of quantum theories.}  

Returning back to the non-unitary evolution caused by singularity inside the black hole, we have similar picture in mind; in the semi-classical regime, the gravity plays a role similar to an external force, while singularity resembles an external device. Singularity can cause jump/collapse of states, resulting to non-unitariness. Although the jump behaves in a non-local way, but it is expected not to have any measurable effect including faster-than-light signalling. This section is provided to investigate this issue.

For an outside observer there is not any jump, \emph{i.e.} evaporation of $B$ is a local unitary process. So, let's focus on infalling observer inside the black hole.  Here we investigate whether the jump of $B$ from singularity to outside of the black hole affects his physical observations. To this end, it would suffice to show that as far as observations are concerned, ``destruction at the singularity" and ``disappearing there and jumping outside the black hole" are indistinguishable.  Classically there is not any possibility for causal contact between him and the jumped $B$ during its evolution afterwards, because of \textbf{(II)}. Hence, classically the issue is simply confirmed. But additional investigation is needed at the quantum level, where non-local correlations can exist. What we do below is to provide a suitable setup for this investigation. Here is a review of a famous and illuminating thought experiment, useful for our later discussion.

\color{black}\paragraph{``Alice and Bob" experiment:} Assume that Alice and Bob have shared two entangled spins. Once Alice performs a measurement on her spin, the state of the whole system collapses to a new disentangled state. But Bob can not distinguish whether any measurement has happened, using his local observables. If he wants to know whether Alice has done a measurement or not, he has two\footnote{There is a third way, the non-local measurement on two spins, but it is forbidden by local physics.} possible ways:

\begin{enumerate}
\item receiving some classical signals from Alice (which might include the Alice's spin or new extra spins),
\item performing some local observation on the two spins, which requires having a trip towards Alice's position.
\end{enumerate}
Otherwise he is not able to extract any information about the collapse, using local measurements on his spin.

\newpage
\paragraph{Non-measurability of the jump at quantum level:}  As a simple but generic enough investigation for the claim, let's modify the Alice and Bob experiment as follows: assume Alice to be close to the singularity of a black hole and Bob be a freely falling observer \emph{inside} the event horizon, but far from the singularity. Alice's spin would disappear in the singularity, while Bob is falling freely inside the black hole. The evolution of the Hilbert space describing the quantum system would be non-unitary, but Bob can not detect any observable effect. It is because there is not any way to perform one of two actions mentioned above; (1) inside a black hole with space-like singularity\footnote{We will not study the black holes with time-like singularities here, because of the presence of multiple or degenerate horizons. They can be dealt in probable extensions of the subject.} (for example the Schwarzschild black hole), there is no possibility for sending any classical signal from a region close to the singularity, to the regions far from it. The causal structure and the direction of time prevents it. (2) Bob can not have a journey towards Alice position unless himself would fall into the singularity where the meaning of observer breaks down. 

Concerning Alice's particle that has jumped outside the black hole, again, none of the two mentioned ways are possible for Bob, in order to detect some physical implications of the non-unitary evolution; (1) According to \textbf{(II)}, the particle jumps in such a way that can not be returned back into the black hole. As a result there is not possibility for sending classical signals from it to  Bob. (2) Bob is inside of black hole, so in order to have a journey to the jumped $B$, he needs to escape the black hole which is impossible. 

To summarize, for infalling observers inside, as far as observations are concerned, the \emph{destruction of particles at singularity} compared to \emph{disappearance of particles at singularity and jumping outside of black hole non-locally} are equivalent and indistinguishable.

\section{A Hot Explosion at the End of Evaporation}\label{sec hot explosion}
As a by-product of our analysis, there is an unexpected but interesting outcome; about half of the entropy of the black hole (carried by inner partners) evaporate just after the Page time $t_{_\text{P}}$ and in a ``short" time interval $\Delta t_\epsilon$ which ends by complete evaporation  of the black hole. In this section we elaborate the logical steps leading to this result.

At first, it is necessary to justify the following role of the Page time, \emph{i.e.} the time when black hole has evaporated half of its entropy:
\begin{quote}
\emph{Page time is the time when the black hole evaporation process switches from usual pair production at the horizon to the quantum jump evaporation.}
\end{quote}
To this end, consider the initial entropy of the black hole to be $S_0$. By definition, at the Page time  the radiated particles (calling them early radiations) and the remaining black hole, each one would possess entropy equal to $\frac{S_0}{2}$. Hence, the radiations which evaporate after Page time (calling them late radiations) would carry the remained $\frac{S_0}{2}$ entropy in the black hole. In brief, each set of the early and late radiations carry half of the initial black hole entropy. This equality of entropy leads to the equality of average number of particles in early and late radiations, which will be described in a moment. This equality of average number of particles justify the claim that Page time is the time when evaporation switches from the evaporation of exterior partners to the jumps of interior partners; because interior and exterior partners are equal in numbers.

To show the equality of average number of particles in the early and late radiations, we can assume the Hawking radiation to be composed of photons for simplicity. According to the elementary thermodynamics of black body radiation which has been just radiated from a black object at temperature $T$, that radiation can be modelled\footnote{Notice that if that black body object is the black hole, this modelling works better, because of centrifugal potential barrier.} by a canonical ensemble of photon gas at temperature $T$, enclosed in some volume $\Delta V$. For such an ensemble, the entropy and mean number of photons can be calculated to be 
\begin{equation}
\Delta S=\frac{4\pi^2\text{k}_{_\mathrm B}^4}{45c^3\hbar^3}T^3 \Delta V \,, \qquad \Delta N=\frac{2\text{k}_{_\mathrm B}^3 \zeta(3)}{\pi^2 c^3\hbar^3}T^3\Delta V \,,
\end{equation}
in which the speed of light $c$, Boltzmann constant $\text{k}_{_\mathrm B}$, Planck constant $\hbar$ and Riemann zeta function $\zeta(n)$ should be understood.  Interestingly,
\begin{equation}\label{S-N const}
\frac{\Delta S}{\Delta N}=\frac{2\pi^4 \text{k}_{_\mathrm B}}{45 \zeta(3)}=\text{const.}\,,
\end{equation}
hence independent of temperature and volume. Now, we can model the part of Hawking radiation in the vicinity of stretched horizon as such an ensemble at some (slowly varying but irrelevant) Hawking temperature. Assuming constancy of $\Delta S$ and $\Delta N$ in that part of Hawking radiation as it propagates outward, then by extensivity of $S$ and $N$ we can deduce that the set of early radiation would have the ratio $\frac{S}{N}$ equal to the constant in \eqref{S-N const}. The same result would be true for the late radiations. Therefore, we reach to the desired result: early and late radiations have equal entropy, and hence equal mean number of photons.

The results above could be expected, because by the AMPS argument the pair creation has to finish before the Page time. On the other hand, jumps can happen only after interior partners reach to the singularity, which ``can not happen so early".  In what follows, we make this latter more precise. 

\begin{figure}[t]
\centering
\definecolor{light gray}{RGB}{220,220,220}
\captionsetup{width=.8\textwidth}	
\begin{tikzpicture}[scale = 3.5] 
  \begin{scope}[scale=1]
   \draw[thick](-0.75,0)--(-0.625,0.125)--(-0.5,0)--(-0.375,0.125)--(-0.25,0)--(-0.125,0.125) -- (0,0)-- (0.125,0.125)-- (0.25,0)-- (0.375,0.125)-- (0.5,0)-- (0.625,0.125)-- (0.75,0)-- (0.875,0.125)-- (1,0) -- (1,1);
  \end{scope};
  \fill[even odd rule,light gray](1,0) to [out=40,in=184] (3,1) to (3,-1) to (2,-1) to node[midway,below,black]{\footnotesize $\Sigma_2$} (1,0);
  \draw (1,0) to [out=40,in=184] node[midway,above,black]{\footnotesize $\Sigma_1$} (3,1);
  \draw (2.9,1) node[anchor=north, black]{\footnotesize $t_\text{end}$};
  \draw[loosely dashed,thick] (0,-1) node[anchor=east]{\footnotesize Event horizon} to (1,0);
  \draw[dashed] (0.3,-1) node[anchor=west]{\footnotesize Stretched horizon} to (1,0);
  \draw (1,0) node [anchor=east,xshift=-0.1cm] {$k$};
  \draw (0.07,-1) to [out=44,in=182] (3,0.7);
  \draw (0.13,-1) to [out=42,in=180](3,0.32);
  \draw (2.9,0.32) node[anchor=north, black]{\footnotesize $t_\epsilon$};
  \draw (0.2,-1) to [out=40,in=180] (3,-0.1);
  \draw [decorate,decoration={brace,amplitude=3pt},xshift=0pt,yshift=0pt]
(1.19,-0.19) -- (0.6,-0.6) node [black,midway,below,xshift=0.2cm,yshift=0.1cm] 
{\footnotesize $\epsilon$};
\end{tikzpicture}

\caption{\footnotesize The conditions \textbf{(I)}, \textbf{(II)} and complementarity confine the allowed region for late evaporation to the gray region, at spatial distance equal or less than $\epsilon$ from the stretched horizon, hence at or after the time $t_\epsilon$. }
\label{fig:epsilon}
\end{figure}
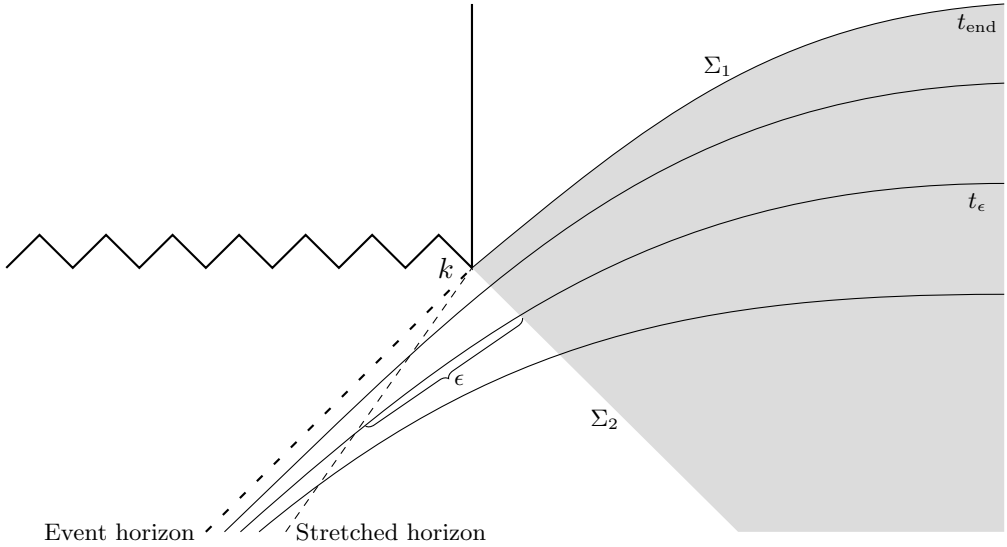

According to \textbf{(I)} and \textbf{(II)} in Section \ref{Our proposal}, the jumps are allowed to happen below $\Sigma_1$ and above $\Sigma_2$ in Figure \ref{fig non-unitary frefo 2}, \emph{i.e.} the gray region. Then, the complementarity of observables outside horizon, explained in the condition \textbf{(III)}, constraints more that region. From the point of view of an outside observer, jumped particles have been created almost on the stretched horizon, \emph{i.e.} an infinitesimal spatial distance (\emph{i.e.} less or equal to Planck length) $\epsilon$ from the stretched horizon. On the other hand, for an infalling observer inside the black hole those particles are created outside by the quantum jump from singularity inside. In a distance far enough the horizon, (roughly at spatial distance equal to $\epsilon$ from the stretched horizon) the two complementary pictures should coincide. So, the jumping particles has to re-appear outside but at $\epsilon$ spatial distance from the stretched horizon, in order to be identical to particles evaporated from stretched horizon as seen by outside observers.  This spatial constraint leads to a bound on the time when jumps can begin as seen by distant observer. The analysis is as follows. The stretched horizon has to disappear when evaporation completes. Hence, it would be a timelike hypersurface ending at the point $k$. Considering this issue plus noticing the allowed gray region, smallness of $\epsilon$ forces us to focus on the spacetime around $k$ in Figure \ref{fig non-unitary frefo 2}. It is depicted in Figure \ref{fig:epsilon}. Some surfaces of constant time are depicted in it, which have to intersect the stretched horizon as the time approaches the end of evaporation. Beginning from the end of evaporation which is labelled by $\Sigma_1$ going backward in time, there would be a certain time $t_\epsilon$ on which the spatial distance from stretched horizon to the gray region boundary (labelled by $\Sigma_2$) would become comparable to $\epsilon$. The times before $t_\epsilon$ are not legitimate for jumps to happen, because the spatial distance between stretched horizon and the gray region gets greater than $\epsilon$.    

Concluding the discussion above, quantum jumps are allowed to happen in a period of time $\Delta t_\epsilon\equiv t_{\text{end}}-t_\epsilon$. It is clear that in the limit $\epsilon \to 0$ (compared to the Planck length), $\Delta t_\epsilon$ tends to zero. Therefore, jumps would be allowed to occur in a shorter and shorter period of time before $t_{\text{end}}$, the time of complete evaporation. The above reasoning delimits the spacetime where the particle can jump into, to the gray region, an infinitesimal spatial distance $\epsilon$ outside of stretched horizon, and in the time interval $\Delta t_\epsilon$ before complete evaporation. An acute reader might have realized that during the time interval between $t_{_\mathrm P}$ and $t_\epsilon$ there is not possibility for any radiation. So, in order to prevent having a black hole without Hawking radiation in that period of time, one can make the identification $t_{_\mathrm P}=t_\epsilon$. As a result, equipped with $t_{_\mathrm P}$ and $\epsilon$, the end of complete evaporation $t_{\text{end}}$ can be determined.

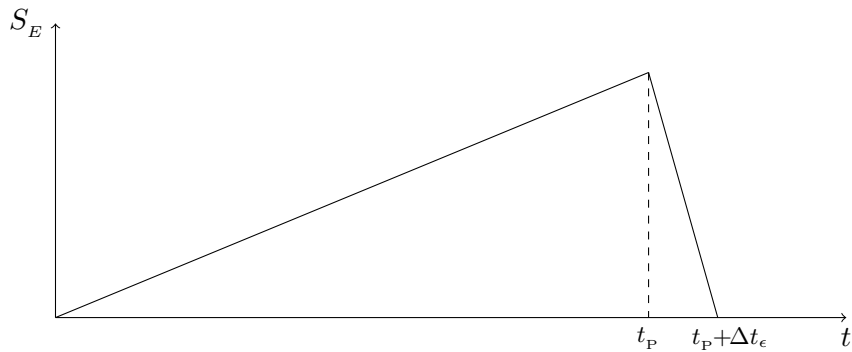
\begin{figure}[t!]
\centering
\vspace*{0.2cm}
\captionsetup{width=.8\textwidth}	
\begin{tikzpicture}[scale = 1.3]
    \draw[->] (0,0) to (0,3) node [anchor = east] {$S_{_E}$};
    \draw[->] (0,0) to (8,0) node [anchor = north] {$t$}; 
    \draw (0,0) to (6,2.5);
    \draw (6,2.5) to (6.7,0)node [anchor = north] {\footnotesize $\,\,\,\,\,\,t_{_\mathrm P}\!\!+\!\!\Delta t_\epsilon$};
    \draw[dashed] (6,2.5) to (6,0) node [anchor = north] {\footnotesize $t_{_\mathrm P}$};
\end{tikzpicture}

\caption{\footnotesize Entanglement entropy between black hole and its Hawking radiation, denoted by $S_{_E}$, increases as time passes through usual pair productions at horizon. It increases monotonically until the Page time $t_{_\mathrm{P}}$. After that, late radiations evaporate in a short period of time $\Delta t_\epsilon$, complement to the jumping process. Leakage of interior partners through jumping process reduces the $S_E$ till the end of evaporation when there would not be any black hole, hence $S_E$ would vanish. As a by-product of this analysis, roughly half of the black hole entropy is released in that short period of time $\Delta t_\epsilon$.}
\label{fig:timeline}
\end{figure}

Let us summarize the timeline of evaporation of the black hole: from the beginning till the Page time, there would be usual Hawking radiation generated by pair production process. After Page time, late radiations jump outside in a relatively short time interval $\Delta t_\epsilon$. If true, late radiations, which because of pair creation process constitute roughly half of the entropy of the black hole, would evaporate as a very hot gas because of high Hawking temperature at the end of black hole evaporation. Besides, lifetime of usual black holes (which their radius is much bigger than Planck length) can be approximated by their Page time $t_{_\mathrm P}$. For completeness, the entanglement entropy between black hole and its Hawking radiation is depicted in Figure \ref{fig:timeline}.

\section{Conclusion and Outlook}

In the semi-classical level, the picture which our approach draws would be unitary evolution for all outside observers (including free-falling observers), and non-unitary evolution for the observers inside the black hole. A transition from unitarity to non-unitarity occurs when a free-falling observer passes the event horizon. It is because the outside observers do not have the singularity as a part of their spacetime, but the inside free-falling observers have. The non-unitary evolution is a non-local behaviour of the Hilbert space, but would not have any measurable effects (and in this sense, it is similar to the collapse of the wave function). Taking the transition into account, the firewall paradox is resolved. Freely falling observers would not encounter any firewall. The resolution can be put in few sentences:
\begin{quote}
\emph{Thanks to the non-unitarity inside black hole, for a free-fall observer, to have a vacuum compatible with the equivalence principle, the late radiations do not need to have any interior partners. In this scenario they are themselves, interior partners of the early radiation. This partnership/entanglement is in agreement with, and what is expected from  the  unitary evolution for the outside observers.}
\end{quote}

Although by the modified complementarity the information and firewall paradox can be resolved, but there would remain so many interesting questions. Among them, two important questions are: 1) is it possible to show dynamically that information is not lost at the horizon, as seen by an outside observer? 2) is it possible to reproduce black hole entropy by studying (thermo)dynamics of evaporated gas? Recently, there have been interesting progresses towards answering these questions. In \cite{Hawking:2015qqa}, S.W. Hawking proposes how information can be stored in the conserved charges associated to supertranslations. In parallel, G. 't Hooft has augmented his proposal for retrieving information through classical back reactions \cite{'tHooft:1984re,'tHooft:1995ij,Hooft:2015jea}. For precursor and recent continuation of these works, \cite{Polchinski:2015cea, Mersini-Houghton:2015yxx,Hawking:2016msc} can be referred to. The progress in this line of research emphasizes convergence and deep connection between the resolutions.

Some other interesting questions towards the future may be as follows: What are probable implications of the postulate {1$'$} for the observers residing outside of black hole? Is it possible to quantify the  timeline of evaporation of black hole in terms of its mass? Can the hot explosion at the end of evaporation of black holes be considered as a source of high energy bursts observed in the sky? Is it possible to make the postulate {1$'$} manifest for other different backgrounds? Does the non-locality in the non-unitary evolution have some conceptual consequences?  How deep can the non-unitarity be? Is it just a property of the semi-classical approximation? Can it help us towards understanding quantum gravity?
\paragraph{Note:} In the period of submission process of this paper, a paper with somehow similar idea has appeared by Lowe and Thorlacius \cite{Lowe:2015eba}. Their idea, which is ``considering the black hole singularity as an erasure of information for a freely falling observer to resolve firewall paradox", is a part of our picture. Nonetheless, the re-appearance of inside modes as late radiations is absent in that work.

\section*{Acknowledgement}

F. S. D. would like to thank Institute for Research in Fundamental Sciences (IPM) for the great hospitality shown during the initial stages of this work, especially to Prof. Sheikh-Jabbari. K. H. would like to thank the Quantum Gravity Group at the  IPM, for useful discussions.  Also he thanks  Raphael Bousso, Vahid Karimipour, Jalaledin Yousefi Koupaei and Leonard Susskind for helpful comments and discussions. He would like to thank Shahin Sheikh-Jabbari, for scientific and financial supports. This work has been supported by the \emph{Allameh Tabatabaii} Prize Grant of \emph{National Elites Foundation} of Iran. We would also thank unknown referee, helping improvements in the presentation.

\appendix

\section{Another Version of the Firewall Paradox}

In this appendix, we tell another version of the firewall paradox, based on the Bogoliubov transformations. After that, we will describe that how non-unitary evolutions in the Hilbert space of freely falling observers, affects the Bogoliubov transformations of their vacuum state.

It is known that the Hawking radiation arises because of the difference between the vacua of the field theory before and after the collapse: Minkowski and Boulware vacua. Minkowski vacuum is seen to consist of particles relative to Boulware vacuum which is the vacuum of stationary observers, i.e. the ones who remain at fixed $r$ coordinate. 

According to a stationary observer who is far from the black hole, the state which the Hawking radiation is in reads as follows \cite{Dundar:2014gma}:
\begin{equation}
  \ket \psi = \alpha \cdot \exp\left(\sum_{lm}\int d\omega \; e^{-4\pi M \omega}a_{\omega lm}^{\text{int} ,\dagger} a_{\omega lm}^{\text{out},\dagger}\right) \ket{\text{B}} \label{HawkingState}
\end{equation}
where $\alpha$ is a complex number and $\ket{\text{B}}$ denotes the Boulware vacuum. This observer remains fixed at some constant but large $r$ coordinate.

If we consider the state of the Hawking radiation seen by other observers which remain at fixed but smaller $r$ we see that the \emph{structure} of the state in equation \eqref{HawkingState} remains the same, although the particles have relatively higher energies because of the blue shift. Suppose $r$ is very close to $2M$ so that the observer is just outside the event horizon.

\begin{figure}[h]
\centering
\captionsetup{width=.8\textwidth}	
\begin{tikzpicture}[scale = 1.5] % 1.4142 = sqrt(2)
  \draw[dashed] (-2,-2) -- (2, 2) node [near end, above, sloped] {Event horizon};
  \draw[dashed] (-2, 2) -- (2,-2);
  \draw[->] (-2,0) -- (2,0) node [anchor=west] {$\rho$};
  \draw[->] (0,-2) -- (0,2) node [anchor=south] {$\tau$};
  
  \draw[thin, domain=-0.5:1.5] plot (\x, {1-2*\x});
  
  \draw[thin, domain=-1.31695789692:1.31695789692] plot ({1.0*cosh(\x)}, {1.0*sinh(\x)});
\end{tikzpicture}

\caption{\footnotesize Approximating the region around the event horizon through Minkowski spacetime: $\tau$ and $\rho$ are temporal and radial Riemann coordinates. Event horizon becomes one of the Rindler horizons. The stationary observer who remains at fixed $r$ becomes a Rindler observer in this description. The straight line is a worldline for a freely falling observer.}
\label{fig:Rindler}
\end{figure}
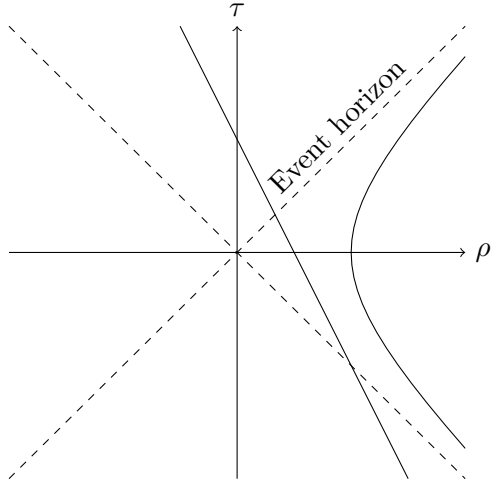

The region of spacetime around the event horizon can be {locally} approximated to great accuracy by flat spacetime. What is interesting now is that the observer who remains at fixed $r$ becomes a Rindler observer in the flat spacetime approximation, see Figure \ref{fig:Rindler}. We have an accelerated motion, and the Hawking radiation detected by the observer is just the Unruh radiation \cite{Unruh:1976db}. Therefore, the Hawking radiation and Unruh radiation share the same entanglement \emph{structure}. Hence, the local vacuum state around the event horizon is directly in correspondence with the specific state of the Hawking radiation which is highly entangled. The transformations which manifest this correspondence are the well-known Bogoliubov transformations.

If one insists on the purity of the Hawking radiation, then the state in equation (\ref{HawkingState}) turns out to be nowhere near describing its quantum state. Hence the state experienced by freely falling observers near the event horizon is nowhere near the vacuum. The event horizon is replaced by a firewall.

But, if one considers the re-appearance of the states inside the black hole outside of the horizon, then some of the basis vectors of the Hilbert space (corresponding to the inside modes) are identified with other basis vectors (corresponding to the outside modes). Hence, the Hilbert space of freely falling observer inside the black hole would lack a complete basis. As a result, the Bogoliubov transformations would not remain valid, and the argument above would break down. Hence, the absence of $A$ modes would not be prohibited. 

\newpage

\bibliographystyle{unsrt}

\end{document}